\documentstyle[aps,preprint,prc]{revtex}

\begin{document}

\begin{center}
{\bf \Large New calculations of the PNC Matrix Element for the
$J^{\pi}T$ 0$^{+}1,0^{-}1$ doublet in $^{14}$N\\}
\vspace{.6cm}
{\bf Mihai Horoi\footnote{permanent address: Department of Theoretical
Physics, Institute of Atomic Physics, Bucharest Romania} and
 G\"{u}nther Clausnitzer}\\

\vspace{.3cm}

{\it Strahlenzentrum der Justus-Liebig-Universit\"{a}t, D-6300 Giessen,
Germany}\\

\vspace{0.5cm}

{\bf B. Alex Brown}\\

\vspace{.3cm}

{\it National Superconducting Cyclotron Laboratory and\\
Department of Physics and Astronomy, East Lansing, MI 48824}\\

\vspace{0.5cm}

{\bf E. K. Warburton}\\

\vspace{.3cm}

{\it Brookhaven National Laboratory, Upton, New York 11973}\\
\end{center}

%\newpage
\vspace{.3cm}

\begin{abstract}
A new calculation of the predominantly isoscalar PNC matrix
element between the $J^{\pi}T$ $0^{+}1,0^{-}1$ (E$_{x} \approx$ 8.7 MeV)
states in $^{14}$N has been carried out in a (0+1+2+3+4)$\hbar
\omega$ model space with the Warburton-Brown interaction.
The magnitude of the PNC matrix element of
0.22 to 0.34 eV obtained
with the DDH PNC interaction is
substantially suppressed
compared with previous calculations in smaller model spaces
but shows agreement with the preliminary Seattle experimental data.
The calculated sign is opposite to that obtained experimentally,
and the implications of this are discussed.

\vspace*{.7cm}


\end{abstract}

\newpage

Studies of low-energy
parity nonconservation (PNC) in light nuclei
have been developed to provide more reliable results on the
hadron-meson weak-coupling constants which are of
importance for our understanding of the quark behavior in nucleons
under the influence of the fundamental interactions. These
studies necessitate both very delicate experiments and very
reliable nuclear structure calculations of the matrix elements
for a correct extraction of the weak nucleon-meson coupling constants.

Most of the results on the experimental and theoretical PNC studies
in light nuclei have been
presented in the
review by Adelberger and Haxton \cite{AH}. From the proposed cases during
the last 25 years in light nuclei, four cases are thought to
reliable for quantitative experimental and theoretical
analysis. They involve parity-mixed doublets (PMD) \cite{AH} in
$^{14}$N, $^{18}$F, $^{19}$F and $^{21}$Ne. Two others cases
involving PMD's in $^{16}$O \cite{KHDC} and $^{20}$F \cite{HC92}
have been  proposed recently. From the four mentioned cases, only
that of $^{19}$F has been measured with a result larger then the
experimental error. The other cases have been measured with errors larger
($^{18}$F and $^{21}$Ne) or near the result ($^{14}$N). However, the
absolute values of the measured errors for $^{18}$F and $^{21}$Ne
are so small that they impose severe constraints on the different
contributions to the PNC matrix elements. These constraints combined with
theoretical calculations indicate a discrepancy,
which has not yet been completely resolved. Namely, if one interprets the
small limit of the (experimentally) extracted PNC matrix element
($<0.029$ eV) for $^{21}$Ne as a destructive interference between
the isoscalar and isovector contributions \cite{AH}, then it
is difficult to understand why the isovector contribution in
$^{18}$F is so small ($<0.09$ eV) and the isoscalar + isovector
contribution in $^{19}$F is relatively so large (0.40 $\pm$ 0.10).
The possibility of an amplification of the isovector contribution
in $^{21}$Ne is not supported by the actual structure
calculations\cite{AH}. However, recent investigations \cite{HCBW93}
indicate that, in the $^{21}$Ne case, the isoscalar contribution
is very small (if not zero) and this could provide an explanation.

Another  possibility for resolving this problem
is to better study the
isoscalar and isovector components separately. Continuous
theoretical and experimental efforts have been undertaken in this
direction. The only case  predominantly isoscalar (no isovector
contribution) is the J$^{\pi}$,T $0^{+}1,0^{-}1$ doublet (E$_{x} \approx 8.7$
MeV) in $^{14}$N. Study of this doublet via the
$^{13}C(\vec{p},p)^{13}C$ resonance scattering was proposed in
1984 \cite{AHB}
and preliminary experimental results were presented in Refs.
\cite{ZP89,ZTh}. The theoretical description of the scattering process is
under control\cite{AHB,HPBCN14} and has been successfully tested
for the regular observables. The predominantly isoscalar PNC
matrix element (the isotensor part contributes $\sim$ 7\%) has been
calculated many times in different model spaces and with
different Hamiltonians. The results vary from 1.39 eV in the
ZBM
($0p_{1/2}$,$0d_{5/2}$,$1s_{1/2}$)
model space\cite{AHB} to 0.29 eV in a full $0p-1s0d$ \ (0+1+2)$\hbar
\omega$ model space using the Kuo-Brown interaction\cite{JH88}.

The aim of this paper is to provide a new analysis of the PNC
matrix element in $^{14}$N based on a new Hamiltonian recently obtained
by Warburton and Brown \cite{WB92} and including also
3$\hbar \omega$ and 4$\hbar \omega$ configurations.
This analysis is of importance for the support of the PNC
experiments in
$^{14}$N
\cite{ZTh,HPBCN14} and to better
understand how to improve the Hamiltonians for a more reliable
description of the weak observables in light nuclei (A = 10-22).

In the last few years, important progress have been made in
the improvement of
the shell-model calculations with special emphasis on the
description of the weak
observables \cite{HJ90,WBM92}. Recently, two new interactions have
been developed by Warburton and Brown \cite{WB92}, which were
designed to
describe pure $\hbar \omega$ states in nuclei with A=10-22.
For this purpose, all of the
0$\hbar \omega$
and 1$\hbar \omega$ and
two-body matrix
elements for the
$0p-1s0d$ model space
have been determined from a least
squares fit to experimental binding energies.
The 1s and 1f2p major shells were also included.
For $^{16}$O and $^{14}$N, (0+2+4)$\hbar \omega$ calculations are now
possible. In the first calculations for $^{16}$O
\cite{HJ90} it was found that a reduction (about 3 MeV)
in the gap between the $0p$ and $1s0d$ major shells is necessary
in order to account to for the
spectrum of
$^{16}$O.
More recently \cite{WBM92}, it has been shown that
this reduction compensates for the
absent
$\geq$ 6$\hbar \omega$ configurations. That is, as is well known,
the effective interaction
and effective single-particle energies are model space
dependent.
Several
choices for the 2$\hbar \omega$
two-body matrix elements within the
$0p-1s0d$ model space have been proposed, based upon the
structure of the mixed (0+2+4)$\hbar \omega$ states in
$^{16}$O
\cite{WBM92,BRO93}.

In order to investigate the sensitivity to various aspects
of the truncation and interaction, we have carried out the PNC calculation for
$^{14}$N
using wave functions obtained with a variety of
assumptions.
Our shell-model calculations have been
performed with the shell-model code OXBASH \cite{OXBA}. Spurious
center of mass motion is removed by the usual method
\cite{GL} of adding a center of mass Hamiltonian to the
interaction. The first four major shells do not provide a
completely nonspurious shell-model basis when more than
2$\hbar \omega$
configurations are included. However, the effect of this spuriosity has
been found it to be negligible.

For the first calculation discussed here we have used
(0+2)$\hbar\omega$
configurations for the positive-parity states
and
(1+3)$\hbar\omega$ configurations for the negative-parity states.
The WBT interaction from Ref.
\cite{WB92} was used for all two-body matrix elements.
In order to reproduce the energy level
spectrum of
$^{14}$N (see Fig. 1),
the $0d_{5/2}$ single-particle energy (SPE) has been lowered by 1 MeV, the
$0p_{1/2}$
SPE increased by 2.2 MeV and the $0p_{3/2}$ SPE
increased by 0.7 MeV. These changes gives a very good spectrum for the
$0^{\pm}$ and $1^{\pm}$ states in $^{14}$N (see Fig. 1)
 and keeps the $\vec{l}
\cdot \vec{s}$ splitting of the $0p$ states at a reasonable value.
(It is recognized that
these changes of the single-particle energies are perhaps arbitrary
and not a unique method
for reproducing the energy spectrum. However,
below we will introduce other models and interactions.)

The PNC matrix element has been calculated in a one-body
approximation. This method was pioneered by Michel \cite{Mi64},
and recently justified and often used for the PNC calculations
\cite{AH,KJV92}. In this paper we have not used the one-body PNC
potential derived in the Fermi gas model approximation (see Eqs.
17-20 of Ref. \cite{AH}). Instead, we have used an exact
calculation of the one-body contribution to the PNC matrix
element

\[ < J^{\pi} T \mid V_{PNC} \mid J^{- \pi} T^{'} >_{OB} =
 \sum_{n_{1}l_{1}j_{1},n_{2}l_{2}j_{2},t} \frac{
C^{T'\ \ t\ T}_{M'\ \tau\ M}}{\sqrt{(2J+1)(2T+1)}}
 \]
\begin{equation}
\times <n_{1}l_{1}j_{1}\parallel
U_{s.p.}^{(0t)} \parallel n_{2}l_{2}j_{2}> $ OBTD $(
(n_{1}l_{1}j_{1})(n_{2}l_{2}j_{2});0t)\ \,
% \label{eq:spme}
\end{equation}

\noindent
where OBTD denotes the one-body-transition density and

\begin{equation}
< \alpha \mid U_{s.p.} \mid \beta > = \sum_{\delta \in core} <\alpha
\delta \mid V_{PNC} \mid \beta \delta > - <\alpha
\delta \mid V_{PNC} \mid \delta \beta >\ ,
\end{equation}

\noindent is assumed.
For $^{14}$N, an interpolation between a $^{12}$C core and a
$^{16}$O core has been performed. This method has been
checked by comparing the one-body calculations (OB) with the full
two-body (TB) calculations (see Table 1. $a$ and $c$).
The OB calculations give results with a precision
of 2\%, at least for the components of the $V_{PNC}$ with the
largest contribution to the matrix element, i.e.

\begin{equation}
V_{PNC}^{\Delta T=0} = - g_{\rho} h_{\rho}^{\circ} (1+\mu_{v})
 \vec{\tau}_{1}
\cdot \vec{\tau}_{2} \imath (\vec{\sigma}_{1} \times
\vec{\sigma}_{2}) \biggl\lbrack
\frac{\vec{p}}{2 M} , \frac{\exp (-m_{\rho}r)}{4
\pi r} \biggl\rbrack \ .
\end{equation}
in the isoscalar case (see Ref. \cite{AH} for notation).

The PNC matrix elements calculated with weak-coupling constants
from different quark models (see Ref. \cite{KHDC} for
notation and references) are presented in Table 1($a$).
The matrix elements were obtained using harmonic-oscillator
wave functions with $\hbar \omega = 14.0$ MeV. (Below we will
address the important issues associated with loosely-bound wave functions.)
The
isotensor contribution has been calculated in the full TB approximation
and is found to give  about a 7\% destructive contribution
to the isoscalar matrix element. In all calculations discussed
below this contribution has been added to the OB result.
Our value of 0.489 eV
obtained in the (0+1+2+3)$\hbar \omega$
model space with the WBT interaction and
based on the DDH best values for the weak couplings, is to
be compared with previous results of 1.39 eV from the
ZBM
($0p_{1/2}$,$0d_{5/2}$,$1s_{1/2}$)
model space with the REWIL interaction \cite{AHB}, 1.04 eV from the
(0+1+2)$\hbar \omega$ model space \cite{DH80},
0.56 eV from the (0+1+2)$\hbar \omega$ model space and the
Millener-Kurath interaction \cite{JH88},
and 0.29 eV from the (0+1+2)$\hbar \omega$ model space with the
Kuo-Brown interaction \cite{JH88}.
The experimental
data \cite{ZP89,ZTh,ZA94}
suggest a magnitude of about $0.38\pm0.28$~eV for the PNC
matrix element.
For comparison, a (1+2)$\hbar \omega$
calculation with the WBT interaction has been performed. The
(0$^{+}1)_{2}$ state in the PMD has been assumed to be a pure
2$\hbar \omega$ configuration, while the (0$^{-}1)_{1}$ state has
been considered a pure 1$\hbar \omega$ configuration. The
calculated binding energies (no modification of the SPE) are quite close
(-107.72 and -108.82 MeV for the $0^{+}1$ and $0^{-}1$ states respectively)
and the PNC matrix element was found to be 1.24 eV for the DDH
weak couplings.

It is interesting to analyze the contributions to the PNC matrix
element in order to understand the source of change when
going to larger model spaces and
the correlations with other calculated observables (e.g. the
electromagnetic transition probabilities). It is convenient to
rewrite the PNC matrix element, Eq. (1), in the following form

\begin{equation}
< (0^{-}1)_{1} \mid V_{PNC}^{\Delta T = 0} \mid (0^{+}1)_{2} >
 = \sum_{\alpha \beta}
\psi_{\alpha \beta} \cal{V}_{\alpha \beta} \equiv \sum_{\alpha
\beta} \cal{C}_{\alpha \beta}
\label{eq:pncs}
\end{equation}

\noindent
where $\alpha ,\ \beta$ denotes the single-particle orbitals,
$\psi_{\alpha \beta}$ is the one-body transition density (OBTD in
Eq. (1) \cite{OXBA}) and
$\cal{V}_{\alpha \beta}$ the single-particle matrix element of
the one-body PNC potential (including the spin-isospin
coefficient in front). The detailed contributions entering Eq.
(\ref{eq:pncs}) are presented in Table 2 for the main component of
V$^{\Delta T=0}_{PNC}$ [ Eq. (3)]. The $\cal{C}_{\alpha
\beta}$ and $\cal{V}_{\alpha \beta}$ are in MeV and they are given
up to an dimensionless factor $-g_{\rho} h_{\rho}^{\circ} (1+
\mu_{v})/2$, which depends on the quark model.
A general analysis of the important  contributions to PNC matrix
element has been made in Ref. \cite{AH}. Our specific
results for
$^{14}$N
are: a.) the main contribution comes from the
$(\alpha \beta)=(0p_{1/2} 1s_{/2})$ amplitude in all model
spaces; this is the only contribution in the ZBM model space; b.)
 a spreading of the strength appears  going to a larger model space,
the $\psi_{0p_{1/2} 1s_{1/2}}$ decreases, the effect of the other
$(\alpha \beta)$ contributions is not more than 20 \% and it is
constructive in all cases (the destructive contribution coming
from $\psi_{0p_{3/2} 0d_{3/2}}$ is rather small); c.) the effect
of the pairing forces in the destructive contribution
$\psi_{\beta \alpha}$ (to $\psi_{\alpha \beta}$) can be seen only
in the ZBM
($0p_{1/2}$,$0d_{5/2}$,$1s_{1/2}$)
and $0p-1s0d$ (0+1+2+3)$\hbar \omega$ model spaces; it contributes
 20\% in the ZBM space and 40\% in the larger
(0+1+2+3)$\hbar \omega$ model space.

It was suggested in Ref. \cite{AH} that the E1 operators could
behave in a similar way with respect to these
 destructive effects due to the fact that they are also odd
under particle-hole conjugation; therefore they
might represent a good test for the reliability of
the wave functions with respect to the axial-charge matrix elements. In Table
3  some electromagnetic transition probabilities
between $0^{\pm}$ and $1^{\pm}$ states in $^{14}$N are included
and compared
with the recent experimental results. The B(M1) are very close to
the experimental results of Zeps \cite{ZP89}. If one excludes the very small
B(E1) transition $(1^{-}0)_{1} \rightarrow (0^{+}1)_{1}$ all
other B(E1) transitions are underestimated in the new calculations
by a factor of 3 to 5; this means a factor of 1.7 to 2.3 for
the matrix element. Can we conclude that
the PNC matrix element is underestimated by a similar factor? In
order to address this question, it is important to look to the
components of the E1 matrix elements and compare with those of
the PNC matrix element. This analysis is presented in Table 4,
for the transition $(0^{+}1)_{2} \rightarrow (1^{-}0)_{1}$, where
we have rewritten the E1 matrix element in the following form

\begin{equation}
<(1^{-}0)_{1} \mid E1 \mid (0^{+}1)_{2} > \ = \ \sum_{\alpha \beta}
\psi_{\alpha \beta} \cal{E}_{\alpha \beta} \equiv \sum_{\alpha
\beta} \cal{B}_{\alpha \beta} \ ,
\end{equation}
similar to Eq. (4). The $\psi_{0p_{1/2} 1s_{1/2}}$ is still one
of the main components, but the other important one comes from
$\psi_{0p_{3/2} 0d_{5/2}}$. Even if the $\psi_{0p_{3/2} 0d_{5/2}}$
admixture is relatively small, its contribution is rather large due to a
large single-particle matrix element, $\cal{E}_{\alpha \beta}$.
 The effect of $\psi_{\beta
\alpha}$ is small and even constructive for the $\psi_{0p_{1/2} 1s_{1/2}}$
component.
So, the origin of the smallness of the E1 matrix element is due to
the cancelation between $\psi_{0p_{1/2} 1s_{1/2}}$ and
$\psi_{0p_{3/2} 0d_{5/2}}$ contributions which does not exist in
the PNC case. One can conclude that in this case the smallness of
the E1 matrix elements does not necessarily indicate an underestimation of
the PNC matrix element.

Another important way to analyse the PNC matrix element
is to consider different
$n\hbar \omega \rightarrow (n\pm 1)\hbar \omega$
contributions. For such an analyses
we have carried out calculations with various strong Hamiltonians and
different methods to treat the effect of the higher $n\hbar \omega$
configurations. We have performed seven different calculations (see
also the code labels in Table 1):\\

\noindent
$a$ - The WBT interaction \cite{WB92} with the SPE modified as discussed above
and with (0+1+2+3)$\hbar \omega$
\  configurations included.\\

\noindent
$b$ - Same as $a$ except that 4$\hbar \omega$
\ configurations are also
included for the $0^{+}1$
states.\\

\noindent
$c$ - The WBT interaction with a modified
$0p-1s0d$ gap ($\Delta \epsilon_{0p} =
0.9 $ MeV, $\Delta \epsilon_{1s0d} = -1.1 $ MeV) and with
(0+1+2+3)$\hbar \omega$
\ configurations included.\\

\noindent
$d$ - Same as $c$ except that 4$\hbar \omega$
\ configurations are also
included for the $0^{+}1$ states.\\

\noindent
$e$ - The WBP interaction \cite{WB92} with
shifted energies \cite{WBM92} ($\Delta 2\hbar \omega = -6.0$ MeV,
$\Delta 1\hbar \omega = -1.75 $ MeV, $\Delta 3\hbar \omega =
-7.25 $ MeV) and with (0+1+2+3)$\hbar \omega$
\ configurations included.\\

\noindent
$f$ - Same as $e$ except that the Bonn potential multiplied by 0.8 has
been used for 2$\hbar \omega$
\ 0p-1s0d cross-shell matrix elements \cite{BRO93}
($\Delta 2\hbar \omega = -6. $ MeV,
$\Delta 1\hbar \omega = -2.5 $ MeV, $\Delta 3\hbar \omega =
-8.5 $ MeV).\\

\noindent
$g$ - The Millener-Kurath interaction \cite{MK75}
with shifted energies ($\Delta 2\hbar \omega = -6.0 $ MeV,
$\Delta 1\hbar \omega = -2.5 $ MeV, $\Delta 3\hbar \omega =
-8.5 $ MeV)
and with (0+1+2+3)$\hbar \omega$
\ configurations.\\

\noindent
The calculated PNC matrix elements are presented in Table 1. The
corresponding spectra and the decompositions in
$n\hbar \omega \rightarrow (n\pm 1)\hbar \omega$
contributions for some of these cases
are presented in Figs. 1-6.
The range of values for the DDH weak-coupling
constants vary between 0.232 and 0.764 eV, with a average
value of around 0.48 eV.

Table 5 presents the relative contributions
to the wave functions of the $(0^{+}1)_{2}$
and $(0^{-}1)_{1}$ states coming from different n$\hbar \omega$
\ configurations.
Table 6 presents the different
$n\hbar \omega \rightarrow (n\pm 1)\hbar \omega$
contributions to the PNC matrix
element (DDH weak couplings assumed). The $(0^{+}1)_{2}$ state is
predominantly a 2$\hbar \omega$
\ configuration. The 0$\hbar \omega$
\ configuration is
small ($<$10\%) but the 0$\hbar \omega$
-1$\hbar \omega$
\ contributions is rather large
and opposite in sign as compared with the dominant 2$\hbar \omega$
-1$\hbar \omega$
\ contribution. This is in fact the main mechanism of suppression
of the PNC matrix element and it is very sensitive to the 0$\hbar \omega$
\ content of the $(0^{+}1)_{2}$ wave function.

Another important point is the sign of the
2$\hbar \omega$ $\rightarrow$ 3$\hbar \omega$
\ contribution. If the $\psi _{0p_{1/2} 1s_{1/2}}$ component would be
dominant for every contribution in the
n$\hbar \omega \rightarrow (n\pm 1)\hbar \omega$
series then the sign
of this contribution should be negative and the PNC matrix
element would be further suppressed. However, all calculations in
Table 6 give a positive sign. This result is
is very sensitive to the mass dependence of
the single-particle energies given by the interaction. All models
in Table 6 correctly describe the experimental SPE order for
$^{13}$C ($1s_{1/2}$ $0d_{5/2}$). Another calculation with a
version of the MK interaction which happened to give an
opposite $0d_{5/2}$-$1s_{1/2}$ SPE order, gives an
opposite sign for the 2$\hbar \omega$ $\rightarrow$ 3$\hbar \omega$
\ contribution and, as a
consequence, a very small PNC matrix element ($\sim$ 0.08 eV).

The (0+2+4)$\hbar \omega$
\ calculations for the $0^{+}1$ states are not
completely fixed. The inclusion of the 4$\hbar \omega$
\ configurations
depresses the $(0^{+}1)_{2}$ state by $\sim$ 3MeV.
(No attempt to
correct for this effect has been made.)
The main
result is that the 4$\hbar \omega$ $\rightarrow$ 3$\hbar \omega$
\ contribution is positive and
smaller than the 2$\hbar \omega$ $\rightarrow$ 3$\hbar \omega$
\ one, suggesting a
convergence of the series. The overall
suppression of the PNC matrix element comes from the
0$\hbar \omega$
\ contribution to the $(0^{+}1)_{2}$ wave function
(9.8\% in case $b$), and thus it is
important to have a good description
of the 0$\hbar \omega$
\ contribution to the $(0^{+}1)_{2}$ wave function.
One can try
to fix this by looking to other observables, e.g. B(E1) and B(M1)
transition probabilities. Different contributions to B(E1) matrix
elements for the $(0^{+}1)_{2} \rightarrow (1^{-}0)_{1}$
transition are presented in Table 8. One can see that the
dominant contributions are
0$\hbar \omega$ $\rightarrow$ 1$\hbar \omega$
and 2$\hbar \omega$
$\rightarrow$ $\hbar \omega$
these are in phase and hence are not very sensitive to the
0$\hbar \omega$
contribution to the $(0^{+}1)_{2}$ wave function.

The B(M1) transitions $(0^{+}1)_{2} \rightarrow (1^{+}0)_{1,2}$
which are, of course, all
n$\hbar \omega$ $\rightarrow$ n$\hbar \omega$
appear to be much more relevant. The various n$\hbar \omega$
contributions to the B(M1) matrix elements are presented in Table
7. The 0$\hbar \omega$
and 2$\hbar \omega$ contributions
are opposite in sign and thus
contribute destructively to the total B(M1). The 0$\hbar \omega$
contributions are not exactly proportional to the total amount of the
0$\hbar \omega$
\ configuration of the $(0^{+}1)_{2}$ wave function.
However, it is
clear that the cases with a relatively higher 0$\hbar \omega$
\ content ($b$,
$d$, $f$, $g$) give a relatively small B(M1) value as compared with
the experimental data in Table 3. From this one may estimate
approximately 3-5\% 0$\hbar \omega$
\ content of the $(0^{+}1)_{2}$ wave
function and
$\sim$ -(0.350 - 0.450) eV for the 0$\hbar \omega$
-1$\hbar \omega$
\ contribution to the PNC
matrix element.

Another ingredient which is often important for the PNC matrix
element is the effect of the derivative operator (see Eq. (3)) on
the tails of the single-particle wave functions (SPWF)
\cite{AH}. The use of Woods-Saxon (WS) SPWF decrease
the matrix element as compared with its value calculated with
harmonic oscillator (HO) SPWF. However, we make two observations
i.) The derivative operator does not act directly on
the SPWF but rather on the matrix element of the short range
form factor, $[\partial /\partial r(exp(-m_{\rho}r)/r)]$.
ii.) It is clear that the use of the WS SPWF
will change the value of the PNC matrix element, especially when
some states in the WS basis are unbound or nearly unbound.

We have estimated the effect of the nearly unbound
$1s_{1/2}$ WS state on the dominant $0p_{1/2} \rightarrow 1s_{1/2}$
contribution to the PNC matrix element. The $1s_{1/2}$ proton
level is unbound by 0.4 MeV in $^{13}$N and is slightly bound by
0.1 MeV in $^{17}$F. For neutrons, the same level is bound by 1.9
MeV in $^{13}$C and by 3.25 MeV in $^{17}$O. We have chosen a
-0.1 MeV value for the $1s_{1/2}$ proton SPE  and -2.0 MeV
for the neutron SPE.
The comparable values for the $0p_{1/2}$ orbit are -2.5 MeV for
protons and -5.0 MeV neutrons. The WS SPWF are obtained by
adjusting the WS well depth to reproduce the above binding energies.
we have found a suppression of the dominant $0p_{1/2} \rightarrow 1s_{1/2}$
contribution to the PNC matrix element of 37\% in the
case of protons and 28\% for neutrons, an average of 32\%.

In the end a few comments concerning the sign of the PNC observable,
the longitudinal analyzing power. The sign found in the experiment
\cite{ZP89,ZTh} is opposite from our calculations as well as from
the initial prediction \cite{AHB}.
The sign of the calculated observable depends on the product  signs
associated with the PNC matrix element  and
the spectroscopic amplitudes,
 which describe the proton decay of the compound states in $^{14}$N.
As in the previous calculation, the dominant
contribution to the PNC matrix element comes from the 1-2$\hbar \omega$
transition (see Figs. 2-3). Moreover, the spectroscopic factors appear to
be stable
quantities for this case. For instance in case (c), we obtained for $C_{0^{+}}$
(see Ref. \cite{AHB} for notations) a value of 0.226/$\sqrt{2}$ - to be
compared with 0.299/$\sqrt{2}$ in ZBM case \cite{AHB} - and 0.977/$\sqrt{2}$
for $C_{0^{-}}$ (to be compared with 1/$\sqrt{2}$ for ZBM). The
$n \hbar \omega$
decompositions of the spectroscopic factors gives 0.210/$\sqrt{2}$ for
the $^{14}N(0^{+}1)_{2} 0 \hbar \omega \rightarrow ^{13}C\ g.s.\ 0 \hbar
\omega$, 0.016/$\sqrt{2}$ for
$^{14}N(0^{+}1)_{2}\ 2 \hbar \omega \rightarrow ^{13}C\ g.s.\ 2 \hbar \omega$,
0.827/$\sqrt{2}$ for
$^{14}N(0^{-}1)_{2}\ 1 \hbar \omega \rightarrow ^{13}C\ g.s.\ 0 \hbar \omega$
and 0.150/$\sqrt{2}$ for the
$^{14}N(0^{-}1)_{2}\ 1 \hbar \omega \rightarrow ^{13}C\ g.s.\ 2 \hbar \omega$.
 The contributions from different $n \hbar \omega$ transitions are
in phase for both spectroscopic factors so they are more stable quantities
than the PNC matrix element. The widths of the $0^{+},\ 0^{-}$ states
calculated with these spectroscopic factors and the method described
in Ref. \cite{AHB} are 5.2 and 1020 keV respectively. They are in
relatively good agreement (if one have in mind that the width is
proportional to the matrix element squared)
with the experimentally extracted ones
\cite{FETAL}: $3.8\pm0.3$ keV and $410\pm20$ keV.

The sign of the isoscalar PNC matrix element has an interesting history.
The earliest calculations based upon the factorization approximation
gave a sign which was consistently opposite to that found experimentally
in the $^{19}$F and $\vec{p}+p$ experiments
\cite{DDH,HO88}. Later,
estimates
of the quark and sum-rule contributions for the nucleon-nucleon
PNC interactions were
added and were found to change the sign of the isoscalar PNC interaction,
bringing the sign in agreement with the experiment
\cite{DDH,HO88}. The actual sign
from the $^{19}$F experiment is not definite because it relies on
calculating the correct sign for a very weak E1 matrix element. Thus we
have at present three pieces of data: (a) $\vec{p}+p$ scattering which
prefers the DDH sign, (b) the $^{19}$F experiment which prefers the DDH
sign but is not certain and (c) the $^{14}$N experiment which prefers
the factorization approximation. It is difficult to reconcile (a) and
(c). Perhaps the reconciliation of (a) and (c) will require "in medium"
modification of the isoscalar PNC weak coupling constants, but further
and more accurate calculations and
experiments will be needed to clarify this puzzle.

In conclusion, new calculations of the predominantly isoscalar
PNC matrix element between the $(0^{+}1)_{2}, (0^{-}1)_{1}$
 states in $^{14}$N
have been performed in a (0+1+2+3+4)$\hbar \omega$ model space using
new Hamiltonians. A new method to calculate the PNC matrix
elements in an one-body approximation has been proposed and shown
to give reliable results as compared with the full two-body
calculations; this method proves to be very useful for
calculations in larger model space, e.g. (0+1+2+3+4)$\hbar \omega$.
The most reasonable range of values for the PNC matrix element was
found to be 0.22 to 0.34
(a 32\% WS suppression included),
which is in reasonable agreement with a magnitude of about $0.38\pm0.28$~eV
deduced from experiment \cite{ZP89,ZTh,ZA94} (even with the upper limit
given by the error, if one dismiss the experimental result as accurate).
Our range of values are suppressed by a factor of 3-4
with respect to the ZBM
($0p_{1/2}$,$0d_{5/2}$,$1s_{1/2}$)
calculations. This suppression comes
mainly from the decrease of the $\psi_{0p_{1/2} 1s_{1/2}}$
OBTD and from a stronger cancellation due to the
particle-hole conjugate transition densities.
All calculated E1 transition probabilities between $0^{\pm},
1^{\pm}$ states in $^{14}$N are  smaller than the
experimental results but, the mechanism of suppression for the E1
matrix elements is different than for the PNC matrix element.
The analysis of the
$n\hbar \omega \rightarrow (n\pm 1)\hbar \omega$
contributions put in evidence the
importance of the 0$\hbar \omega$
\ content of the
$(0^{+}1)_{2}$ wave function. This part can be to some extent fixed by the
B(M1) transition probabilities.
The relative sign of the
2$\hbar \omega$ $\rightarrow$ 3$\hbar \omega$
\ contributions appear to be fixed as positive but its magnitude
remains as somewhat uncertain. The effect of higher ($>$3$\hbar \omega$
)
configurations deserves further study as well as the convergence
of the
$n\hbar \omega \rightarrow (n\pm 1)\hbar \omega$
series.\\

\vspace*{1.cm}

The authors want to thank Dr. V.J. Zeps for valuable suggestions concerning the
content of the manuscript and for allowing them to use the new experimental
electromagnetic transitions strengths before publication.
M.H. would like to thank the Alexander von Humboldt Foundation for a fellowship
during which this work has been done. B.A.B. would like to acknowledge
support from the Alexander van Humboldt Foundation and NSF grant 90-17077.

%\vspace*{1cm}

%\newpage

\newpage
\vspace{3cm}

\begin{center}
{\bf Figure captions}
\end{center}

\vspace{1.5cm}

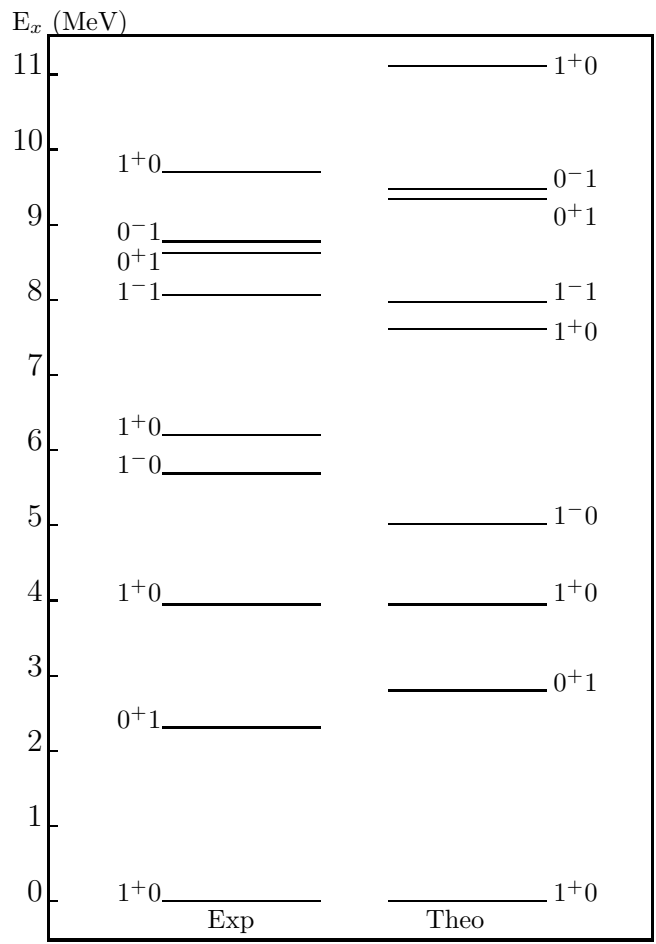
{\bf Figure 1} \ Experimental and calculated
$0^{\pm},1^{\pm}$ levels in $^{14}$N. The calculation is that of
model $a$ obtained with the WBT interaction \cite{WB92}
and modified SPE within the
(0+1+2+3)$\hbar \omega$ model space as described in the text.
\\

{\bf Figure 2} \ Decomposition of the PNC matrix element into the
contributions
coming from different $n\hbar \omega$ components of the wave functions.
The calculation is that of model $a$ and the DDH weak-coupling constants
have been used.
Units are eV.\\

{\bf Figure 3} \ Same as Figure 2 but with model $b$ where
4$\hbar \omega$ configurations are also included.\\

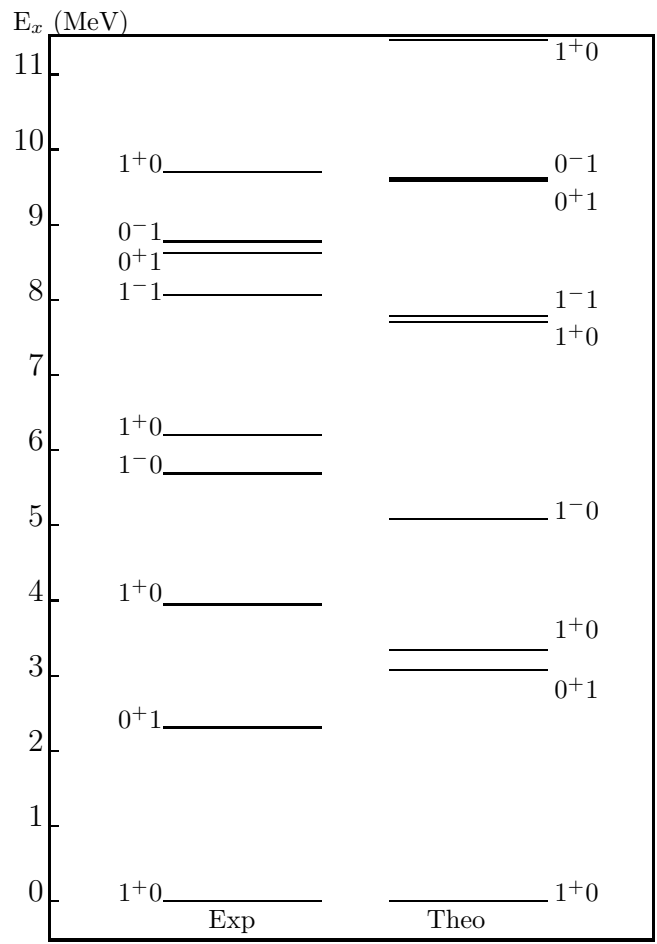
{\bf Figure 4} \ Same as Figure 1 but with model $c$.\\

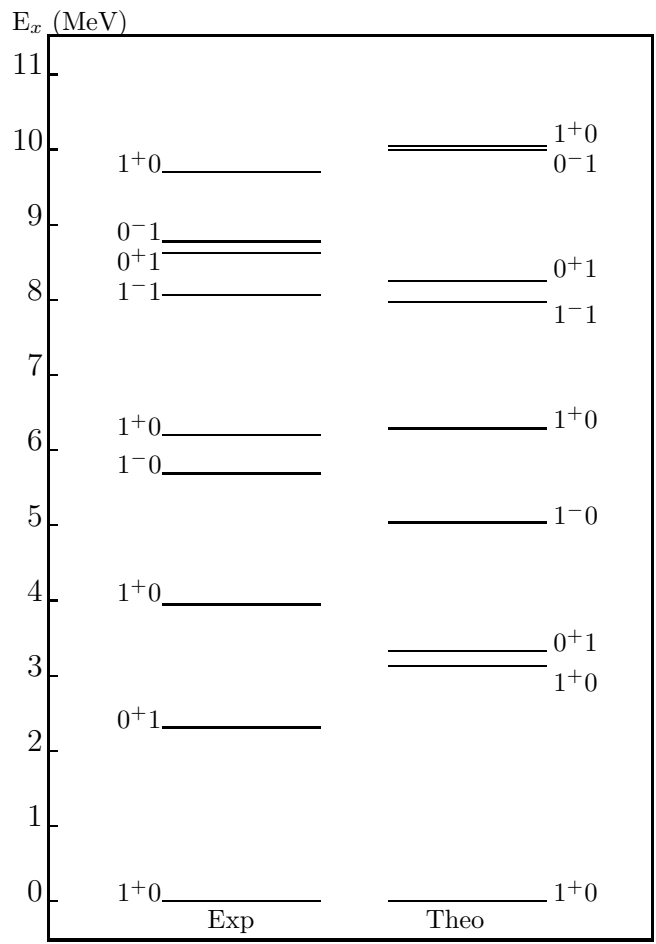
{\bf Figure 5} \ Same as Figure 1 but with model $e$.\\

\newpage

{\bf Table 1} \ Magnitude of the
PNC matrix element calculated
with the various strong interactions \cite{WB92}, different
model spaces, and different models of the
weak-coupling constants (see Ref. \cite{KHDC} for
the weak interaction notation and
references). Units are eV.
Code labels are further explained in the text.
\\

\begin{center}

\begin{tabular}{|c|c|c|c||c|c|c|c|}
\hline
& & & & \multicolumn{4}{|c|}{Weak-Coupling Models}\\
\cline{5-8}
Interaction & Code & Model space & PNC & DDH & AH & DZ & KM \\
\hline
\hline
WBT  & $a$ & (0+1+2+3)$\hbar \omega$ & OB &0.487 & 0.372 & 0.421 & 0.278\\
\cline{2-8}
SPE  & $a$ & (0+1+2+3)$\hbar \omega$ & TB &0.483 & 0.366 & 0.413 & 0.269\\
\cline{2-8}
modification & $b$ & (0+1+2+3+4)$\hbar \omega$ & OB &0.233 & 0.164 & 0.190 &
0.1
27\\
\hline
WBT  & $c$ &(0+1+2+3)$\hbar \omega$ & OB &0.764 & 0.565 & 0.620 & 0.418\\
\cline{2-8}
gap & $c$ & (0+1+2+3)$\hbar \omega$ & TB &0.732 & 0.549 & 0.620 & 0.400\\
\cline{2-8}
modification & $d$ & (0+1+2+3+4)$\hbar \omega$ & OB &0.497 & 0.366 & 0.413 &
0.2
72\\
\hline
WBP & $e$ & (0+1+2+3)$\hbar \omega$
& OB &0.502 & 0.370 & 0.418 & 0.275\\
\cline{2-8}
 & $e$ & (0+1+2+3)$\hbar \omega$ & TB &
0.492 & 0.371 & 0.417 & 0.269\\
\hline
WBP + 0.8 Bonn & $f$ & (0+1+2+3)$\hbar \omega$
& OB & 0.351 & 0.260 & 0.292 & 0.193\\
\hline
MK & $g$ & (0+1+2+3)$\hbar \omega$
& OB & 0.331 & 0.246 & 0.276 & 0.181 \\
\hline
\hline
\end{tabular}\\

\end{center}

\newpage

{\bf Table 2} \ Components of the PNC matrix element entering Eq.
(4) as described in  text.
\\

\begin{center}
{\footnotesize
\begin{tabular}{|c|c|c||c|c||c|c||c|c|}
\hline
 & & & \multicolumn{2}{|c||}{ZBM} & \multicolumn{2}{c||}{(1+2)$\hbar
\omega$} & \multicolumn{2}{c|}{(0+1+2+3)$\hbar \omega$} \\ \cline{4-9}
$\alpha$ & $\beta$ & $\cal{V}_{\alpha \beta}$
& $\psi_{\alpha \beta}$ & $\cal{C}_{\alpha \beta}$
& $\psi_{\alpha \beta}$ & $\cal{C}_{\alpha \beta}$
& $\psi_{\alpha \beta}$ & $\cal{C}_{\alpha \beta}$\\
\hline
\hline
$0p_{1/2}$ & $1s_{1/2}$ & 0.171 & 1.193 & 0.204 & 0.717 &
0.1226 & 0.510 & 0.0871\\
$1s_{1/2}$ & $0p_{1/2}$ & -0.171 & 0.259 & -0.044 &  &
& 0.235 & -0.0402\\
\hline
$1s_{1/2}$ & $0p_{1/2}$ & -0.196 & & & -0.082 &
0.0161 & -0.045 & 0.00881\\
$0p_{1/2}$ & $1s_{1/2}$ & 0.196 & &  & &
 & -0.0006 & -0.0001\\
\hline
$0p_{3/2}$ & $0d_{3/2}$ & -0.213 & &  & 0.004 &
-0.0009 & 0.044 & -0.0102\\
$0d_{3/2}$ & $0p_{3/2}$ & 0.213 & & &  &
 & 0.028 & 0.0063\\
\hline
$1s_{1/2}$ & $2p_{1/2}$ & -0.169 & & & 0.072 &
0.0122 & -0.015 & 0.0025\\
$2p_{1/2}$ & $1s_{1/2}$ & 0.169 &  & &  &
 & 0.015 & 0.0025\\
\hline
\hline
\end{tabular}}\\

\end{center}

\newpage

{\bf Table 3} \ Experimental and calculated electromagnetic
transition probabilities, B(M1) and B(E1), between $0^{\pm},
1^{\pm}$ states in $^{14}$N. Units are $\mu_N^{2}$ and $e^{2}
fm^{2}$, respectively.\\

\begin{center}
{\footnotesize
\begin{tabular}{|cc|c||c|c||c|c|}
\hline
& & & \multicolumn{2}{|c||}{Experiment} & \multicolumn{2}{c|}{Theory}\\
\cline{4-7}
\multicolumn{2}{|c|}{transition}
 & type & AjS \cite{AjS} & Zeps \cite{ZA94} & ZBM &
(0+1+2+3)$\hbar \omega$ \\
\hline
\hline
$(0^{+}1)_{2}$ & $\rightarrow (1^{+}0)_{1}$ & M1 & 0.159 &
0.05$\pm$0.005 & 0.32 & 0.031\\
 & $\rightarrow (1^{+}0)_{2}$ & M1 & 1.056 &
1.05$\pm$0.1 &  & 0.572\\
 & $\rightarrow (1^{+}0)_{3}$ & M1 & 12.71 &
12.2$\pm$1.2 & 12.7 & 11.31\\
 & $\rightarrow (1^{-}0)_{1}$ & E1 & 0.0258 &
0.0161$\pm$0.0019 & 0.16 & 0.0042\\
\hline
$(0^{-}1)_{1}$ & $\rightarrow (1^{+}0)_{1}$ & E1 & 0.0636$\pm$0.0187 &
0.0355$\pm$0.0028 & 0.086 & 0.015\\
\hline
$(1^{-}0)_{1}$ & $\rightarrow (0^{+}1)_{1}$ & E1 &
((0.42$\pm0.19)\cdot 10^{-3})$ & & & $0.44 \cdot 10^{-2}$\\
\hline
$(1^{-}1)_{1}$ & $\rightarrow (1^{+}0)_{1}$ & E1 &
(1.8$\pm0.45)\cdot 10^{-2}$ & (1.23$\pm0.09)\cdot 10^{-2}$ &
 & $0.62 \cdot 10^{-2}$\\
 & $\rightarrow (1^{+}0)_{2}$ & E1 &
(2.1$\pm0.45)\cdot 10^{-2}$ & (1.47$\pm0.12)\cdot 10^{-2}$ &
 & $0.46 \cdot 10^{-2}$\\
\hline
\hline
\end{tabular}}\\

\end{center}

\newpage

{\bf Table 4} \ Components of the E1 matrix element entering Eq.
(5) for the $(0^{+}1)_{2} \rightarrow (1^{-}0)_{1}$ transition.\\

\begin{center}
{\footnotesize
\begin{tabular}{|c|c|c||c|c||c|c|}
\hline
 & & & \multicolumn{2}{|c||}{ZBM} &
 \multicolumn{2}{c|}{(0+1+2+3)$\hbar \omega$} \\ \cline{4-7}
$\alpha$ & $\beta$ & $\cal{E}_{\alpha \beta}$
& $\psi_{\alpha \beta}$ & $\cal{B}_{\alpha \beta}$
& $\psi_{\alpha \beta}$ & $\cal{B}_{\alpha \beta}$\\
\hline
\hline
$0p_{1/2}$ & $1s_{1/2}$ & 0.485 & 0.6886 & 0.334 & 0.2804 & 0.136 \\
$1s_{1/2}$ & $0p_{1/2}$ & 0.485 & 0.1493 & 0.072 & 0.1191 & 0.057\\
\hline
$0p_{3/2}$ & $0d_{5/2}$ & -1.455 & & & 0.0514 & -0.0749 \\
$0d_{5/2}$ & $0p_{3/2}$ & 1.455 & &  & 0.0029 & 0.00415 \\
\hline
\hline
\end{tabular}}\\

\end{center}

\newpage

{\bf Table 5} Relative contributions of the $n\hbar \omega$ excitations to
the wave functions of $(0^{+}1)_{2}$ and $(0^{-}1)_{1}$ states in
$^{14}$N. Code labels correspond to the cases in Table 1.\\

\begin{center}
\begin{tabular}{|c|c||c|c|c|c|c|}
\hline
\hline
Code & $J^{\pi}T$ & 0$\hbar \omega$
 & 1$\hbar \omega$
 & 2$\hbar \omega$
 & 3$\hbar \omega$
 & 4$\hbar \omega$
 \\
\hline
\hline
& $0^{+}1$ & \ \ 0.047\ \  & & \ \ 0.953\ \  & & \ \ \ \  \\
%$a$ & & & & & &\\
$a$ & $0^{-}1$ & & \ \ 0.855\ \  & & \ \ 0.145\ \  & \\
\hline
& $0^{+}1$ & \ \ 0.098\ \  & & \ \ 0.773\ \  & & \ \ 0.129\ \  \\
%$b$ & & & & & &\\
$b$ & $0^{-}1$ & & \ \ 0.855\ \  & & \ \ 0.145\ \  & \\
\hline
& $0^{+}1$ & \ \ 0.031\ \  & & \ \ 0.969\ \  & & \ \ \ \  \\
%$c$ & & & & & &\\
$c$ & $0^{-}1$ & & \ \ 0.838\ \  & & \ \ 0.162\ \  & \\
\hline
& $0^{+}1$ & \ \ 0.073\ \  & & \ \ 0.763\ \  & & \ \ 0.164\ \  \\
%$d$ & & & & & &\\
$d$ & $0^{-}1$ & & \ \ 0.838\ \  & & \ \ 0.162\ \  & \\
\hline
& $0^{+}1$ & \ \ 0.061\ \  & & \ \ 0.939\ \  & & \ \ \ \  \\
%$e$ & & & & & &\\
$e$ & $0^{-}1$ & & \ \ 0.818\ \  & & \ \ 0.182\ \  & \\
\hline
& $0^{+}1$ & \ \ 0.078\ \  & & \ \ 0.922\ \  & & \ \ \ \  \\
%$f$ & & & & & &\\
$f$ & $0^{-}1$ & & \ \ 0.782\ \  & & \ \ 0.218\ \  & \\
\hline
& $0^{+}1$ & \ \ 0.050\ \  & & \ \ 0.950\ \  & & \ \ \ \  \\
%$g$ & & & & & &\\
$g$ & $0^{-}1$ & & \ \ 0.876\ \  & & \ \ 0.124\ \  & \\
\hline
\hline
\end{tabular}\\

\end{center}

\newpage

{\bf Table 6} The $n\hbar \omega \rightarrow (n\pm 1)\hbar
\omega$ contributions to the PNC matrix element (DDH weak couplings
assumed) for various cases studies (described by code labels).
Units are eV.\\

\begin{center}
\begin{tabular}{|c||c|c|c|c|}
\hline
\hline
Code  & 0$\hbar \omega$
-1$\hbar \omega$
& 2$\hbar \omega$
$\rightarrow$
1$\hbar \omega$
 &
 2$\hbar \omega$
$\rightarrow$
3$\hbar \omega$
&
3$\hbar \omega$
$\rightarrow$
4$\hbar \omega$
\\
\hline
\hline
$a$ & -0.440 & 0.793 & 0.172 &  \\
$b$ & -0.638 & 0.732 & 0.121 & 0.079 \\
$c$ & -0.347 & 0.941 & 0.228 &  \\
$d$ & -0.538 & 0.804 & 0.152 & 0.113 \\
$e$ & -0.453 & 0.764 & 0.222 &  \\
$f$ & -0.530 & 0.717 & 0.190 &  \\
$g$ & -0.344 & 0.669 & 0.032 &  \\
\hline
\hline
\end{tabular}\\

\end{center}

\newpage

{\bf Table 7} $n\hbar \omega \rightarrow n\hbar \omega$
contributions to the relevant M1 matrix elements
in units of $\mu_N$.
Last column contains the calculated B(M1) ($\mu_N^{2}$). Relative
contributions of $n\hbar \omega$ excitations to the wave
functions are also given.\\

\begin{center}

\begin{tabular}{|c|c||c|c||c|c||c|c|c|c|}
\hline
 & & \multicolumn{2}{|c|}{$(0^{+}1)_{2}$} &
 \multicolumn{2}{|c|}{$(1^{+}0)_{n}$} &
 \multicolumn{3}{|c|}{M1 matrix element} & B(M1) \\
\cline{3-10}
Transition & Code & 0$\hbar \omega$
 & 2$\hbar \omega$
 & 0$\hbar \omega$
 & 2$\hbar \omega$
 & 0$\hbar \omega$
 & 2$\hbar \omega$
 & total & \\
\hline
\hline
$(0^{+}1)_{2}$
& $a$ & 0.047 & 0.953 &
0.765 & 0.235 &
-0.108 & 0.284 & 0.176 & 0.031 \\
$\rightarrow$
$(1^{+})_{1}$
 & $b$ & 0.098 & 0.773 & 0.765 & 0.235 & -0.160 & 0.270 & 0.110 & 0.012 \\
 & $c$ & 0.031 & 0.969 & 0.747 & 0.253 & -0.162 & 0.452 & 0.290 & 0.084 \\
 & $d$ & 0.073 & 0.763 & 0.747 & 0.253 & -0.263 & 0.382 & 0.119 & 0.014 \\
 & $e$ & 0.061 & 0.939 & 0.719 & 0.281 & -0.274 & 0.546 & 0.273 & 0.074 \\
 & $f$ & 0.078 & 0.922 & 0.735 & 0.265 & -0.198 & 0.250 & 0.051 & 0.0026 \\
 & $g$ & 0.050 & 0.950 & 0.809 & 0.191 & -0.140 & 0.238 & 0.108 & 0.012 \\
\hline
$(0^{+}1)_{2}$
& $a$ & 0.047 & 0.953 &
0.676 & 0.324 &
-0.485 & 1.242 & 0.757 & 0.572 \\
$\rightarrow$
$(1^{+})_{2}$
 & $b$ & 0.098 & 0.773 & 0.676 & 0.324 & -0.160 & 0.270 & 0.110 & 0.012 \\
 & $c$ & 0.031 & 0.969 & 0.698 & 0.302 & -0.375 & 0.999 & 0.625 & 0.390 \\
 & $d$ & 0.073 & 0.763 & 0.698 & 0.302 & -0.263 & 0.382 & 0.119 & 0.014 \\
 & $e$ & 0.061 & 0.939 & 0.651 & 0.349 & -0.488 & 1.192 & 0.704 & 0.495 \\
 & $f$ & 0.078 & 0.922 & 0.610 & 0.390 & -0.552 & 1.252 & 0.707 & 0.490 \\
 & $g$ & 0.050 & 0.950 & 0.727 & 0.273 & -0.476 & 0.852 & 0.376 & 0.141 \\
\hline
\hline
\end{tabular}\\

\end{center}

\newpage
\oddsidemargin-0.5cm
\evensidemargin-0.5cm

{\bf Table 8} $n\hbar \omega \rightarrow (n\pm 1)\hbar \omega$
contributions to the $(0^{1})_{2} \rightarrow (1^{-}0)_{1}$ E1
matrix elements in units $e fm$.
Last column contains the calculated B(E1) ($e^{2} fm^{2}$). Relative
contributions of $n\hbar \omega$ excitations to the wave
functions are also given.\\

\begin{center}

\begin{tabular}{|c||c|c||c|c||c|c|c|c|c|c|}
\hline
 & \multicolumn{2}{|c|}{$(0^{+}1)_{2}$} &
 \multicolumn{2}{|c|}{$(1^{-}0)_{1}$} &
 \multicolumn{5}{|c|}{E1 matrix element} & B(E1) \\
\cline{2-11}
Code & 0$\hbar \omega$
 & 2$\hbar \omega$
 &  1$\hbar \omega$
 & 3$\hbar \omega$
 &
0
$\rightarrow 1$$\hbar \omega$
& 2
$\rightarrow 1$$\hbar \omega$
& 2
$\rightarrow 3$$\hbar \omega$
 & 4
$\rightarrow 3$$\hbar \omega$
& total
& \\
\hline
\hline
$a$ & 0.047 & 0.953 &  0.839 & 0.161 &
0.026 & 0.053 & -0.014 & & 0.065 & 0.0042 \\
  $b$ & 0.098 & 0.773 & 0.839 & 0.161 & 0.037 & 0.040 & -0.014 & 0.0001
 & 0.063 & 0.0040 \\
  $c$ & 0.031 & 0.969 & 0.828 & 0.172 & 0.015 & 0.055 & -0.017
 & & 0.053 & 0.0029 \\
  $d$ & 0.073 & 0.763 & 0.828 & 0.172 & 0.024 & 0.033 & -0.016 & 0.0014
 & 0.039 & 0.0015 \\
  $e$ & 0.061 & 0.939 & 0.876 & 0.124 & 0.013 & 0.038 & -0.013
  & & 0.039 & 0.0016 \\
  $f$ & 0.078 & 0.922 & 0.778 & 0.222 & 0.016 & 0.040 & -0.006
  & & 0.051 & 0.0026 \\
  $g$ & 0.050 & 0.950 & 0.876 & 0.124 & 0.041 & 0.067 & 0.011
  & & 0.119 & 0.0142 \\
\hline
\hline
\end{tabular}\\

\end{center}

%\vspace{2cm}
\newpage

\topmargin0cm
\textheight23.5cm
\textwidth17cm
\headheight0cm
\headsep0cm

\vspace*{4cm}

\begin{center}

\begin{minipage}[t]{10cm}
\unitlength1.cm

\begin{picture}(8.5,12.5)(0,0)
\thicklines
%\put(0,0){\framebox(16,25)}
\put(.5,.5){\framebox(8,12)}
\thinlines
\put(.0,12.6){\footnotesize E$_{x}$ (MeV)}
%\put(-.25,11.6){\footnotesize MeV}
\put(2.0,1){\line(1,0){2.1}}
\put(2.0,4.948){\line(1,0){2.1}}
\put(2.0,7.204){\line(1,0){2.1}}
\put(2.0,10.703){\line(1,0){2.1}}
\put(2.0,6.691){\line(1,0){2.1}}
\put(2.0,3.314){\line(1,0){2.1}}
\put(2.0,9.620){\line(1,0){2.1}}
\put(2.0,9.776){\line(1,0){2.1}}
\put(2.0,9.062){\line(1,0){2.1}}
\put(5.0,1){\line(1,0){2.1}}
\put(5.0,4.948){\line(1,0){2.1}}
\put(5.0,8.618){\line(1,0){2.1}}
\put(5.0,12.108){\line(1,0){2.1}}
\put(5.0,6.020){\line(1,0){2.1}}
\put(5.0,3.803){\line(1,0){2.1}}
\put(5.0,10.341){\line(1,0){2.1}}
\put(5.0,10.476){\line(1,0){2.1}}
\put(5.0,8.975){\line(1,0){2.1}}
\put(2.6,.65){\footnotesize Exp}
\put(5.5,.65){\footnotesize Theo}
%\put(2.1,2.18){\vector(0,-1){.98}}
%\put(2.,2.06){\vector(0,-1){1.06}}
%\put(5.0,1.69){\vector(0,-1){.69}}
%\put(4.8,2.13){\vector(0,-1){.93}}
%\put(2.1,2.18){\circle*{.05}}
%\put(5.0,1.69){\circle*{.05}}
%\put(2.8,1){\footnotesize 2$^{+}$1}
\put(1.4,9.0){\footnotesize 1$^{-}$1}
\put(1.4,1.0){\footnotesize 1$^{+}$0}
\put(1.4,5.0){\footnotesize 1$^{+}$0}
\put(1.4,7.2){\footnotesize 1$^{+}$0}
\put(1.4,10.7){\footnotesize 1$^{+}$0}
\put(1.4,6.7){\footnotesize 1$^{-}$0}
\put(1.4,3.31){\footnotesize 0$^{+}$1}
\put(1.4,9.4){\footnotesize 0$^{+}$1}
\put(1.4,9.8){\footnotesize 0$^{-}$1}
\put(7.2,9.0){\footnotesize 1$^{-}$1}
\put(7.2,1.0){\footnotesize 1$^{+}$0}
\put(7.2,5.0){\footnotesize 1$^{+}$0}
\put(7.2,8.46){\footnotesize 1$^{+}$0}
\put(7.2,12.0){\footnotesize 1$^{+}$0}
\put(7.2,6.02){\footnotesize 1$^{-}$0}
\put(7.2,3.8){\footnotesize 0$^{+}$1}
\put(7.2,10.0){\footnotesize 0$^{+}$1}
\put(7.2,10.5){\footnotesize 0$^{-}$1}

\put(.5,1){\line(1,0){.1}}
\put(.5,2){\line(1,0){.1}}
\put(.5,3){\line(1,0){.1}}
\put(.5,4){\line(1,0){.1}}
\put(.5,5){\line(1,0){.1}}
\put(.5,6){\line(1,0){.1}}
\put(.5,7){\line(1,0){.1}}
\put(.5,8){\line(1,0){.1}}
\put(.5,9){\line(1,0){.1}}
\put(.5,10){\line(1,0){.1}}
\put(.5,11){\line(1,0){.1}}
\put(.5,12){\line(1,0){.1}}
\put(.2,1){0}
\put(.2,2){1}
\put(.2,3){2}
\put(.2,4){3}
\put(.2,5){4}
\put(.2,6){5}
\put(.2,7){6}
\put(.2,8){7}
\put(.2,9){8}
\put(.2,10){9}
\put(.0,11){10}
\put(.0,12){11}

\end{picture}

\end{minipage}

\begin{center}
Figure 1.
\end{center}

\end{center}

\newpage

%\vspace*{1cm}

\begin{center}

\begin{minipage}[t]{12cm}
\unitlength1.cm

\begin{picture}(10.5,10.5)(0,0)
\thicklines
%\put(0,0){\framebox(16,25)}
%\put(.5,.5){\framebox(10,10)}
%\thinlines
\put(1.4,8.5){\makebox(0,0)[bl]{$\vert ^{14}$N(0$^{+}1)_{2}\ >$\ \
 =\ 0.216$\vert 0\ \hbar \
\omega >$\ \ +\ \ 0.976$\vert 2\ \hbar \ \omega >$}}

\put(1.5,5.){\makebox(0,0)[bl]{$<V_{PNC}^{\Delta T=0}>_{DDH}$
=\ \ \ -0.440\ \ +\ \ 0.793\ \ +\ \ 0.172}}

\put(1.4,1.5){\makebox(0,0)[bl]{$\vert ^{14}$N(0$^{-}1)_{1}\ >$\ \
=\ 0.925$\vert 1\ \hbar \
\omega >$\ \ +\ \ 0.381$\vert 3\ \hbar \ \omega >$}}

\put(5.6,8.2){\line(0,-1){2.8}}
\put(8.9,8.2){\line(-1,-2){1.4}}
\put(8.9,8.2){\line(0,-1){2.8}}

\put(5.6,4.7){\vector(0,-1){2.8}}
\put(7.2,4.7){\vector(-1,-2){1.4}}
\put(8.9,4.7){\vector(0,-1){2.8}}

\end{picture}
\end{minipage}
\begin{center}
Figure 2.
\end{center}
\end{center}

%\vspace*{0.5cm}

\begin{center}

\begin{minipage}[t]{12cm}
\unitlength1.cm

\begin{picture}(10.5,10.5)(0,0)
\thicklines
%\put(0,0){\framebox(16,25)}
%\put(.5,.5){\framebox(10,10)}
%\thinlines
\put(0.5,8.5){\makebox(0,0)[bl]{$\vert ^{14}$N(0$^{+}1)_{2}\ >$\ \
 =\ \ 0.313$\vert 0\ \hbar \
\omega >$\ +\ 0.879$\vert 2\ \hbar \ \omega >$
\ +\ 0.359$\vert 4\ \hbar \ \omega >$}}

\put(0.3,5.){\makebox(0,0)[bl]{$<V_{PNC}^{\Delta T=0}>_{DDH}$
=\ \ \ \ -0.638\ \ +\ \ 0.732\ \ \ \ \ +\ \ \ \ \ 0.121\ \ +\ \ 0.079}}

\put(0.5,1.5){\makebox(0,0)[bl]{$\vert ^{14}$N(0$^{-}1)_{1}\ >$\ \
=\ \ \ 0.925$\vert 1\ \hbar \
\omega >$\ \ \ \ \ \ \ \ \ \ +\ \ \ \
\ \ \ \ \ \ \ 0.381$\vert 3\ \hbar \ \omega >$}}

\put(4.9,8.2){\line(0,-1){2.8}}
\put(7.7,8.2){\line(-1,-3){.94}}
\put(8.3,8.2){\line(1,-3){.94}}
\put(10.9,8.2){\line(0,-1){2.8}}

\put(4.9,4.7){\vector(0,-1){2.8}}
\put(6.46,4.7){\vector(-1,-3){.94}}
\put(9.45,4.7){\vector(1,-3){.94}}
\put(10.9,4.7){\vector(0,-1){2.8}}

\end{picture}

\end{minipage}

\begin{center}
Figure 3.
\end{center}

\end{center}

\newpage

\vspace*{4cm}

\begin{center}

\begin{minipage}[t]{10cm}
\unitlength1.cm

\begin{picture}(8.5,12.5)(0,0)
\thicklines
%\put(0,0){\framebox(16,25)}
\put(.5,.5){\framebox(8,12)}
\thinlines
\put(.0,12.6){\footnotesize E$_{x}$ (MeV)}
%\put(-.25,11.6){\footnotesize MeV}
\put(2.0,1){\line(1,0){2.1}}
\put(2.0,4.948){\line(1,0){2.1}}
\put(2.0,7.204){\line(1,0){2.1}}
\put(2.0,10.703){\line(1,0){2.1}}
\put(2.0,6.691){\line(1,0){2.1}}
\put(2.0,3.314){\line(1,0){2.1}}
\put(2.0,9.620){\line(1,0){2.1}}
\put(2.0,9.776){\line(1,0){2.1}}
\put(2.0,9.062){\line(1,0){2.1}}
\put(5.0,1){\line(1,0){2.1}}
\put(5.0,4.341){\line(1,0){2.1}}
\put(5.0,8.706){\line(1,0){2.1}}
\put(5.0,12.461){\line(1,0){2.1}}
\put(5.0,6.088){\line(1,0){2.1}}
\put(5.0,4.079){\line(1,0){2.1}}
\put(5.0,10.617){\line(1,0){2.1}}
\put(5.0,10.583){\line(1,0){2.1}}
\put(5.0,8.792){\line(1,0){2.1}}
\put(2.6,.65){\footnotesize Exp}
\put(5.5,.65){\footnotesize Theo}
\put(1.4,9.0){\footnotesize 1$^{-}$1}
\put(1.4,1.0){\footnotesize 1$^{+}$0}
\put(1.4,5.0){\footnotesize 1$^{+}$0}
\put(1.4,7.2){\footnotesize 1$^{+}$0}
\put(1.4,10.7){\footnotesize 1$^{+}$0}
\put(1.4,6.7){\footnotesize 1$^{-}$0}
\put(1.4,3.31){\footnotesize 0$^{+}$1}
\put(1.4,9.4){\footnotesize 0$^{+}$1}
\put(1.4,9.8){\footnotesize 0$^{-}$1}
\put(7.2,8.9){\footnotesize 1$^{-}$1}
\put(7.2,1.0){\footnotesize 1$^{+}$0}
\put(7.2,4.5){\footnotesize 1$^{+}$0}
\put(7.2,8.4){\footnotesize 1$^{+}$0}
\put(7.2,12.2){\footnotesize 1$^{+}$0}
\put(7.2,6.09){\footnotesize 1$^{-}$0}
\put(7.2,3.7){\footnotesize 0$^{+}$1}
\put(7.2,10.2){\footnotesize 0$^{+}$1}
\put(7.2,10.7){\footnotesize 0$^{-}$1}

\put(.5,1){\line(1,0){.1}}
\put(.5,2){\line(1,0){.1}}
\put(.5,3){\line(1,0){.1}}
\put(.5,4){\line(1,0){.1}}
\put(.5,5){\line(1,0){.1}}
\put(.5,6){\line(1,0){.1}}
\put(.5,7){\line(1,0){.1}}
\put(.5,8){\line(1,0){.1}}
\put(.5,9){\line(1,0){.1}}
\put(.5,10){\line(1,0){.1}}
\put(.5,11){\line(1,0){.1}}
\put(.5,12){\line(1,0){.1}}
\put(.2,1){0}
\put(.2,2){1}
\put(.2,3){2}
\put(.2,4){3}
\put(.2,5){4}
\put(.2,6){5}
\put(.2,7){6}
\put(.2,8){7}
\put(.2,9){8}
\put(.2,10){9}
\put(.0,11){10}
\put(.0,12){11}

\end{picture}

\end{minipage}

\begin{center}
Figure 4.
\end{center}

\end{center}

\newpage

\vspace*{4cm}

\begin{center}

\begin{minipage}[t]{10cm}
\unitlength1.cm

\begin{picture}(8.5,12.5)(0,0)
\thicklines
%\put(0,0){\framebox(16,25)}
\put(.5,.5){\framebox(8,12)}
\thinlines
\put(.0,12.6){\footnotesize E$_{x}$ (MeV)}
%\put(-.25,11.6){\footnotesize MeV}
\put(2.0,1){\line(1,0){2.1}}
\put(2.0,4.948){\line(1,0){2.1}}
\put(2.0,7.204){\line(1,0){2.1}}
\put(2.0,10.703){\line(1,0){2.1}}
\put(2.0,6.691){\line(1,0){2.1}}
\put(2.0,3.314){\line(1,0){2.1}}
\put(2.0,9.620){\line(1,0){2.1}}
\put(2.0,9.776){\line(1,0){2.1}}
\put(2.0,9.062){\line(1,0){2.1}}
\put(5.0,1){\line(1,0){2.1}}
\put(5.0,4.132){\line(1,0){2.1}}
\put(5.0,7.287){\line(1,0){2.1}}
\put(5.0,10.996){\line(1,0){2.1}}
\put(5.0,6.040){\line(1,0){2.1}}
\put(5.0,4.329){\line(1,0){2.1}}
\put(5.0,9.257){\line(1,0){2.1}}
\put(5.0,11.050){\line(1,0){2.1}}
\put(5.0,8.975){\line(1,0){2.1}}
\put(2.6,.65){\footnotesize Exp}
\put(5.5,.65){\footnotesize Theo}
%\put(2.1,2.18){\vector(0,-1){.98}}
%\put(2.,2.06){\vector(0,-1){1.06}}
%\put(5.0,1.69){\vector(0,-1){.69}}
%\put(4.8,2.13){\vector(0,-1){.93}}
%\put(2.1,2.18){\circle*{.05}}
%\put(5.0,1.69){\circle*{.05}}
%\put(2.8,1){\footnotesize 2$^{+}$1}
\put(1.4,9.0){\footnotesize 1$^{-}$1}
\put(1.4,1.0){\footnotesize 1$^{+}$0}
\put(1.4,5.0){\footnotesize 1$^{+}$0}
\put(1.4,7.2){\footnotesize 1$^{+}$0}
\put(1.4,10.7){\footnotesize 1$^{+}$0}
\put(1.4,6.7){\footnotesize 1$^{-}$0}
\put(1.4,3.31){\footnotesize 0$^{+}$1}
\put(1.4,9.4){\footnotesize 0$^{+}$1}
\put(1.4,9.8){\footnotesize 0$^{-}$1}
\put(7.2,8.7){\footnotesize 1$^{-}$1}
\put(7.2,1.0){\footnotesize 1$^{+}$0}
\put(7.2,3.8){\footnotesize 1$^{+}$0}
\put(7.2,7.3){\footnotesize 1$^{+}$0}
\put(7.2,11.1){\footnotesize 1$^{+}$0}
\put(7.2,6.02){\footnotesize 1$^{-}$0}
\put(7.2,4.33){\footnotesize 0$^{+}$1}
\put(7.2,9.3){\footnotesize 0$^{+}$1}
\put(7.2,10.7){\footnotesize 0$^{-}$1}

\put(.5,1){\line(1,0){.1}}
\put(.5,2){\line(1,0){.1}}
\put(.5,3){\line(1,0){.1}}
\put(.5,4){\line(1,0){.1}}
\put(.5,5){\line(1,0){.1}}
\put(.5,6){\line(1,0){.1}}
\put(.5,7){\line(1,0){.1}}
\put(.5,8){\line(1,0){.1}}
\put(.5,9){\line(1,0){.1}}
\put(.5,10){\line(1,0){.1}}
\put(.5,11){\line(1,0){.1}}
\put(.5,12){\line(1,0){.1}}
\put(.2,1){0}
\put(.2,2){1}
\put(.2,3){2}
\put(.2,4){3}
\put(.2,5){4}
\put(.2,6){5}
\put(.2,7){6}
\put(.2,8){7}
\put(.2,9){8}
\put(.2,10){9}
\put(.0,11){10}
\put(.0,12){11}

\end{picture}

\end{minipage}

\begin{center}
Figure 5.
\end{center}

\end{center}

\end{document}